\documentclass[aps,prl,twocolumn,reprint,groupedaddress,superscriptaddress]{revtex4-1}

\usepackage{graphicx}
\usepackage{amsmath}

\begin{document}


\title{Realization of nonlinear optical nonreciprocity on a few-photon level based on atoms strongly coupled to an asymmetric cavity}

\author{Pengfei Yang}
\affiliation{State Key Laboratory of Quantum Optics and Quantum Optics Devices,
and Institute of Opto-Electronics, Shanxi University, Taiyuan 030006, China}
\affiliation{Collaborative Innovation Center of Extreme Optics, Shanxi University, Taiyuan 030006, China}
\author{Xiuwen Xia}
\affiliation{School of Mathematics and Physics, Jinggangshan University, Ji’an, Jiangxi 343009, China}
\author{Hai He}
\author{Shaokang Li}
\author{Xing Han}
\affiliation{State Key Laboratory of Quantum Optics and Quantum Optics Devices,
and Institute of Opto-Electronics, Shanxi University, Taiyuan 030006, China}
\affiliation{Collaborative Innovation Center of Extreme Optics, Shanxi University, Taiyuan 030006, China}
\author{Peng Zhang}
\affiliation{Department of Physics, Renmin University of China, Beijing 100872, China}
\author{Gang Li}
\email[]{gangli@sxu.edu.cn}
\author{Pengfei Zhang}
\affiliation{State Key Laboratory of Quantum Optics and Quantum Optics Devices,
and Institute of Opto-Electronics, Shanxi University, Taiyuan 030006, China}
\affiliation{Collaborative Innovation Center of Extreme Optics, Shanxi University, Taiyuan 030006, China}
\author{Jinping Xu}
\author{Yaping Yang}
\affiliation{MOE Key Laboratory of Advanced Micro-Structure Materials, School of Physics Science and Engineering, Tongji University, Shanghai 200092, China}
\author{Tiancai Zhang}
\email[]{tczhang@sxu.edu.cn}
\affiliation{State Key Laboratory of Quantum Optics and Quantum Optics Devices,
and Institute of Opto-Electronics, Shanxi University, Taiyuan 030006, China}
\affiliation{Collaborative Innovation Center of Extreme Optics, Shanxi University, Taiyuan 030006, China}

\date{\today}

\begin{abstract}
Optical nonreciprocity is important in photonic information processing to route the optical signal or prevent the reverse flow of noise. By adopting the strong nonlinearity associated with a few atoms in a strongly coupled cavity QED system and an asymmetric cavity configuration, we experimentally demonstrate the nonreciprocal transmission between two counterpropagating light fields with extremely low power. This nonreciprocity can even occur on a few-photon level due to the high optical nonlinearity of the system. The working power can be flexibly tuned by changing the effective number of atoms strongly coupled to the cavity. The idea and result can be applied to optical chips as optical diodes by using fiber-based cavity QED systems. Our work opens up new perspectives for realizing optical nonreciprocity on a few-photon level based on the nonlinearities of atoms strongly coupled to an optical cavity.

\end{abstract}



\maketitle

The phenomenon of optical nonreciprocity (ONR), which allows unidirectional transmission of a light field, always accompanies the physical processes of time-reversal symmetry breaking. Electromagnetic nonreciprocity \cite{Caloz2018} can be harnessed for important devices in information processing systems to route the electromagnetic signal or prevent the reverse flow of noise. In the context of the rapid development of photonic information processing, the realization of ONR, especially nonmagnetic ONR for chip-based optical information processing, has been studied extensively. However, despite the enormous experimental progress in terms of ONR, most studies focused on the control of the light field with classic mechanisms \cite{Hua2016, Jiang2016, DelBino2018, Shadrivov2010, Grigoriev2011,Gallo1999, Gallo2001, Fan2012, Yu2009, Lira2012, Kim2017, Kim2014, Kamal2010, Sounas2013, Estep2014, Chang2014, Peng2014, Shi2015, Bender2013}, and these kinds of ONR cannot alleviate the stringent requirement for extremely low power in chip-based photonic information processing.

In recent years, some novel systems have been experimentally explored to demonstrate the ONR on a few-photon or even single-photon level. The chiral interaction \cite{Lodahl2017} between quantum emitter and whispering-gallery-mode (WGM) microresonators or photonic nanostructures offers new platforms for realizing single-photon-level ONR. Single-photon-level ONR was experimentally observed with this chiral interaction between a single atom and a WGM bottle resonator \cite{Junge2013}. After that, the single-photon-level optical diode and circulator were demonstrated by the chiral interaction between atoms and a nanofiber and WGM bottle resonator \cite{Sayrin2015, Scheucher2016}. Single-photon-level nonreciprocal quantum operations, such as a single photon switch \cite{Shomroni2014} and a photon-atom SWAP gate \cite{Bechler2018}, were also realized on chiral quantum optics systems. There are also other ONR schemes or systems \cite{Shen2014, Lenferink2014, Fratini2014, Fratini2016, Shen2011, XiaK2014, Xu2017, Yan2018, Zhang2018, Wang2013} that have the potential to work on a single-photon or a few-photon level. As was classified in \cite{Caloz2018}, all these ONRs were based on time-reversal symmetry breaking by an external bias in the linear case. In the nonlinear case, the working power of the ONR is very difficult to decrease because a large number of photons are usually involved to obtain observable nonlinearity. Fortunately, the quantum interaction between the light field and material can provide pronounced nonlinearity on the few-quanta level. Some theoretical proposals have noted that few-photon-level ONR is possible by using quantum nonlinearity \cite{Roy2010, Roy2013, Roy2017}, and microwave nonreciprocity resulting from quantum nonlinearity has been demonstrated recently by coupled superconducting qubits \cite{Rosario2018}. In a strongly coupled atom-cavity system, the nonlinear interaction can be observed on the single-photon level \cite{Birnbaum2005, Schuster2008, Hamsen2017}, which thus provides the possibility to realize ONR with extremely low power. 

In this letter, we report an experimental demonstration of ONR on a few-photon level in the nonlinear case, corresponding to the power of $p$W. The nonlinearity comes from a few atoms strongly coupled to an asymmetric optical cavity, which has asymmetric couplings and losses for the two ports. Thus, the ONR can be expected at certain input powers. Benefiting from the strong-coupling-induced high nonlinearity, the ONR can be observed with a few intracavity photons. The blocking ratio, which is defined by the ratio between the cavity transmissions from both sides, for the reversely propagating light field is greater than 15 dB at this working power. We also find that the ONR working window can be tuned by controlling the effective number of atoms in the cavity, and a maximum blocking ratio of 30 dB is reached. The transmission of forward light is approximately 18\%, which is mainly limited by poor impedance matching and extra losses of the cavity. Higher transmission can be achieved by optimizing the impedance matching according to the actual extra-losses under the precondition of an asymmetric cavity. The idea and result we reported here can be easily integrated into optical chips as an optical diode \cite{Jalas2013} by using cavity QED systems with chip-based WGM cavities \cite{Vahala2003, Englund2007, Hennessy2007, Fink2008, Dayan2008, Tiecke2014, Goban2015, Kato2015, Oshea2013} or fiber cavities \cite{Volz2011, Gallego2018}. Our work opens up new perspectives for realizing ONR on a few-photon level based on quantum nonlinearities.

\begin{figure}
\includegraphics[width=8.6cm]{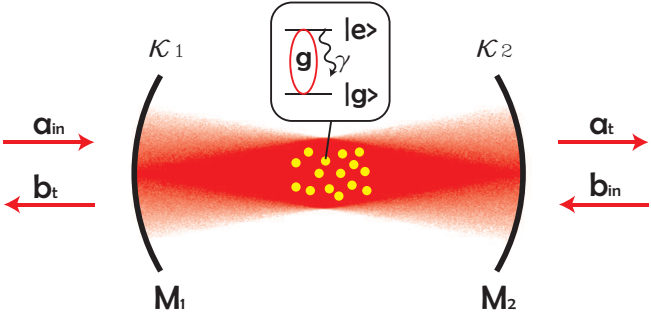}
\caption{\label{fig1} Scheme of the ONR model. A cavity QED system with multiple two-level atoms strongly coupled to an asymmetric cavity is adopted. The coupling efficiencies of the cavity mirrors fulfill the relation $\kappa_1>\kappa_2$, which means that the intracavity atoms reach saturation easier for the incident light field in mode $a$ (the forward direction) than that in mode $b$ (the reverse direction) along with the increase in incident power. Consequently, the light field in mode $a$ can transmit, whereas the light field in mode $b$ is blocked under certain powers. }
\end{figure}

\begin{figure}
\includegraphics[width=8.6cm]{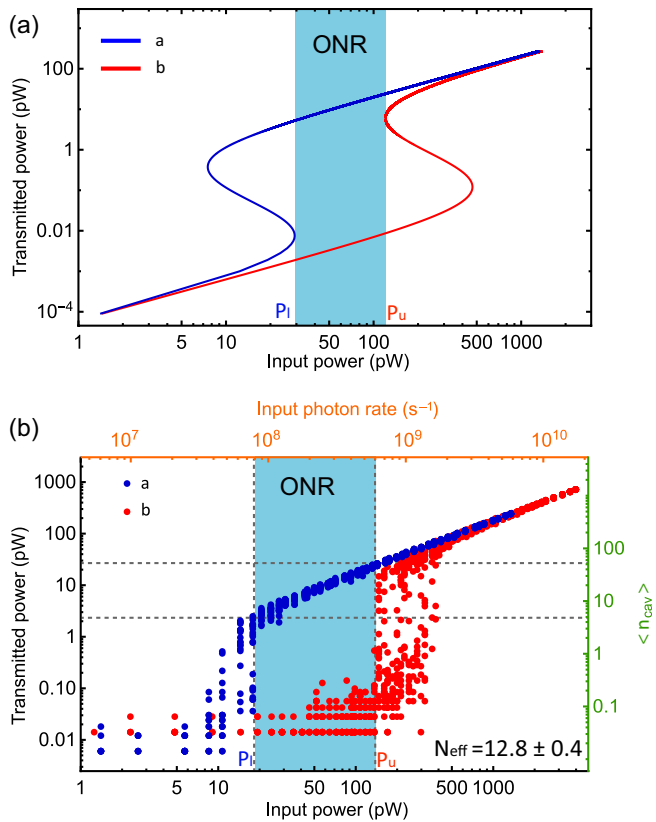}
\caption{\label{fig2} Bistability of the light fields with an effective atom number $N_\text{eff}=12.8 \pm 0.4$ as the power in mode-$a$ and $b$ increases. (a) shows the theoretical results given by the semiclassic method; (b) shows the experimental results. The blue (red) curve and data points are for the forward (backward) light field in mode $a$ ($b$). The shaded area in both figures, with lower and upper bounds marked by $P_\text{l}$ and $P_\text{u}$, indicates the ONR working window, in which the light field in mode $a$ transmits, whereas the light field in mode $b$ is blocked. In (b), the top horizontal and right vertical axes show the input photon rate and the mean photon numbers in the cavity, deduced by the output of the forward light field, respectively.}
\end{figure}

The prototype of the ONR model on which our experiment was based was first developed in \cite{Xia2014}, where a bad cavity with $\gamma < \Omega < \kappa$ was considered, with $\Omega$ being the atom-cavity coupling strength,  and $\gamma$ and $\kappa$ being the atom and cavity decay rates, respectively. Here, we adopt a strongly coupled system, where $\Omega > (\gamma, \ \kappa)$, so that the nonlinearity is much larger than that of the system in weak coupling system. The scheme is shown in Fig. \ref{fig1}. $N$ two-level atoms strongly couple to an optical Fabry-P\'erot cavity. The atom-cavity coupling strength for a single atom is $g$; thus, the collective atom-cavity coupling strength for $N$ atoms is $\Omega=\sqrt{N} g$. The decay rate of the atom from the excited state $|e\rangle$ is $\gamma$, and the cavity decay rate is $\kappa=\kappa_1+\kappa_2 + \kappa_\text{loss}$, where $\kappa_{1(2)}$ is the decay rate (also the coupling rate between photons inside the cavity and outside of the cavity) from mirrors M$_{1(2)}$ and $\kappa_\text{loss}$ is the overall extra-loss-induced decay rate of the cavity mode. The light field with frequency $\omega_\text{p}$ excites the system either from the left ($a_\text{in}$ mode) or right ($b_\text{in}$ mode) side. The frequency detuning between the light field and atomic transition (cavity resonance) is denoted by $\Delta=\omega_\text{at}-\omega_\text{p}$ ($\delta=\omega_\text{cav}-\omega_\text{p}$), where $\omega_\text{at}$ and $\omega_\text{cav}$ are the resonant frequencies of the atomic transition and cavity. After using the standard semi-classical method \cite{Xia2014, Dombi2013} (also refer to the supplementary material, SM \cite{N20}), the relation between the transmitted light power $P^{a(b)}_\text{t}$ ($a$ and $b$ in the superscript mean the directions of the incident field, as shown in Fig. \ref{fig1}) and the incident light power  $P^{a(b)}_\text{in}$ is given by 
\begin{multline}\label{eq1}
P^{a(b)}_\text{in}=\frac{P^{a(b)}_\text{t}}{4 \kappa_{1} \kappa_{2}}\Bigg\{ \left[ \kappa+\frac{\Omega^2 \gamma}{(\Delta^2 + \gamma^2)(1+y)} \right]^2 \\
+\left[ - \delta+\frac{\Omega^2 \Delta}{(\Delta^2 + \gamma^2)(1+y)} \right] ^2 \Bigg\},
\end{multline}
where $y=P^{a(b)}_\text{t}/P^{a(b)}_\text{ct}$ is the saturation parameter, with $P^{a(b)}_\text{ct}=[{\kappa_{2(1)}(\Delta^2+\gamma^2)}]/{g^2}$. In the current research, only the resonant case, i.e., $\Delta \approx \delta \approx 0$, is considered. If we focus only on the light field propagation along one direction, Eq.(\ref{eq1}) gives a bistable behavior between transmitted and incident light fields due to the nonlinearity of the coupled atom-cavity system. This bistability has been extensively studied in either weakly coupled or strongly coupled atom-cavity systems \cite{Grant1982, Rempe1991, Joshi2003, Amitabh2010}. However, for an asymmetric cavity, where $\kappa_1 > \kappa_2 $ for the two mirrors, the nonlinearities for two counterpropagating fields with the same power are quite different due to the asymmetric coupling coefficients $\kappa_1$ and $\kappa_2$. The field incident from the mirror with a larger coupling efficiency has stronger nonlinearity, and the consequential bistability appears for a lower input power. The expected bistability results according to Eq. (\ref{eq1}) for the two light fields are shown in Fig. \ref{fig2}(a). In the shaded region, the atomic transition for the light field in mode $a$ has been saturated, and almost all the light field transmits the cavity, whereas the atomic transition for the light field in mode $b$ is far from the saturation due to the much weaker intracavity field; thus, the light beam is blocked due to normal mode splitting \cite{Thompson1992, Boca2004, Maunz2005}. The expression for the output power of the light field feeding in the $a$- and $b$-modes can be simplified to the first order of the parameter $C/y$ as \cite{N20} 
\begin{equation}\label{eq2}
P^{a}_\text{t}=\frac{4 \kappa_1 \kappa_2}{\kappa^2} P_\text{in}-\frac{2 N \kappa_2 \gamma}{\kappa}
\end{equation}
and 
\begin{equation}\label{eq3}
P^{b}_\text{t}=\frac{4 \kappa_1 \kappa_2}{\kappa^2} \frac{P_\text{in}}{(1+2C)^2},
\end{equation}
where $C=Ng^2/(2\kappa \gamma)$ is the parameter of cooperativity. We can see that the output of mode $a$ is determined not only by the impedance matching of the cavity [the coefficient of the first term on the right-hand side of Eq. (\ref{eq2})] but also by the atomic decay [the second term on the right-hand side of Eq. (\ref{eq2})]. However, the output of mode $b$ is suppressed by the strong coupling, as discussed before, and the stronger the coupling is, the lower the output power. For convenience, we also denote mode $a$($b$) as the forward(backward) direction.

To experimentally demonstrate the ONR, we use a high-finesse Fabry-P\'erot cavity with a length of 335 $\mu$m. The cold cesium atoms are transferred from a magneto-optical trap (MOT) beside the cavity with the aid of a 1064-nm optical dipole trap (see SM \cite{N20} for details). The cavity QED parameters are $(g,\kappa,\gamma)=2\pi \times(5.5, 3.7, 2.6)$ MHz, where $g$ is the maximum atom-cavity coupling strength for the $6S_{1/2} |F=4\rangle \leftrightarrow 6P_{3/2} |F’=5\rangle$ transition of a single atom, and $\kappa$ and $\gamma$ are the field decay rates from the cavity and atom, respectively. The two cavity mirrors $M_1$ and $M_2$ have asymmetric transmission coefficients of $T_1=88.9 \pm 0.1$ ppm and $T_2=5.1 \pm 0.1$ ppm, which correspond to decay (coupling) rates of $\kappa_1=2 \pi \times 3.1$ MHz and $\kappa_2=2 \pi \times 0.2$ MHz, respectively. There is 10.8 ppm extra scattering and absorption losses for the whole cavity, which gives an extra loss rate of $\kappa_\text{loss}=2 \pi \times 0.4$ MHz. A cavity-resonant light field is fed into the cavity from either side of the cavity, and the transmitted light field is recorded by a single photon detector. There is a small detuning between the cavity and the atomic transition, with $\Delta - \delta=(-0.64 \pm 0.2)$ MHz throughout the measurements.

The typical bistability results obtained with an effective intracavity atom number of $N_\text{eff}=12.8 \pm 0.4$ for the two forward and reverse light fields are shown in Fig \ref{fig2}(b). The effective atom number is deduced from the Rabi splitting spectrum in a separate measurement. As expected by the theory, there is an obvious window (the shaded region) in which one light field is transmitted and the other is blocked. This is the ONR working window, and the input working power is between 20 and 140 $p$W for the incident light field. The corresponding intracavity mean photon number of the forward light is from 4.1 to 50. Compared to the theoretical result in Fig. \ref{fig2}(a), the experiment result shows two distinct features. 1) The transmitted power below the onset power of the bistability [see the red dots below 140 $p$W of the input power in Fig. \ref{fig2}(b)] is almost constant despite the input power, while the theory shows dependence with a positive slope. This discrepancy occurs because the transmitted light field is so weak that the background counts of the photon detector dominate the signal. 2) The theoretical bistability from the semiclassic model shows an ``s'' shape in the bistable region, which has been verified by other experiments \cite{Grant1982, Rempe1991, Joshi2003, Amitabh2010}. However, in our experiments, the result shows a noisy distribution in this region. These results come from the quantum process and can be understand by a quantum treatment. For quantum treatment, the two states in the bistable region are no longer stable, and stochastic switching between them occurs continuously \cite{Wu2018}. The full quantum method \cite{Dombi2013} will provide a complete description. In either case, the ONR feature of the light field is still very distinct and holds. 

\begin{figure}
\includegraphics[width=8.6cm]{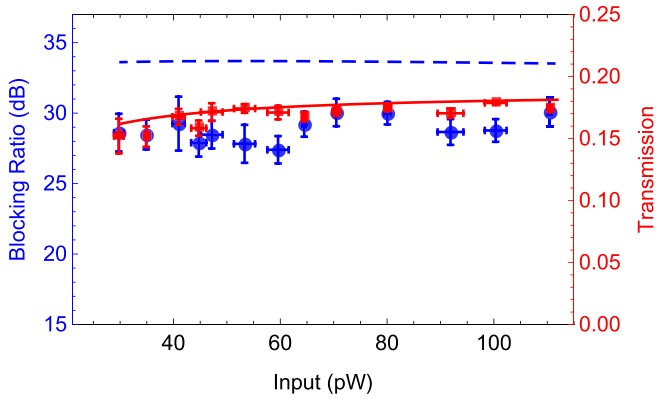}
\caption{\label{fig3} Performance of the ONR with atom number $N_\text{eff}=12.8 \pm 0.4$. The samples of the working power are selected in between 30 and 110 $p$W. The red and blue data points are the experimental results for transmission in mode $a$ and the blocking ratio of the reverse incident light field, respectively. The solid red and dashed blue curves are the theoretical expectations.}
\end{figure}

The performance of the ONR is characterized by transmission efficiency for the forward field and the blocking ratio for the reverse light field. The experimental results are shown in Fig. \ref{fig3}, along with the theoretical expectations. The measured transmission for the forward (mode $a$) light field is approximately 18\%, which is in good agreement with the theory. In the current case with an input power at the $p$W level, the second term in the right part of Eq. (\ref{eq2}) can be eliminated, and the transmission is mainly limited by poor impedance matching, with $\kappa_1>\kappa_2 +\kappa_\text{loss}$. For a given extra loss rate $\kappa_\text{loss}$, higher transmission can be achieved by optimizing the impedance matching, i.e., $\kappa_1=\kappa_2+\kappa_\text{loss}$. In this situation, the mechanism of the nonlinear ONR with an asymmetric cavity ($\kappa_1 > \kappa_2$) can still be fulfilled. For our experimental system, if the total cavity decay rate $\kappa=2 \pi \times 3.7$ MHz and extra cavity loss rate $\kappa_\text{loss}=2 \pi \times 0.4$ MHz remain constant, the highest transmission (78\%) can be achieved by setting $\kappa_1=2 \pi \times 1.85$ MHz and $\kappa_2=2 \pi \times 1.45$ MHz. The blocking ratio is independent of the losses but relies on the parameter of cooperativity, with $P^{a}_\text{t}/P^{b}_\text{t}=(1+2C)^2$ \cite{N20}. The measured average blocking ratio for the reverse light field within the ONR working window is approximately 28 dB, which is less than the theoretical blocking ratio of approximately 34 dB due to the domination of the background counts associated with the single photon detector in the case of an extremely weak probe light field. 

\begin{figure}
\includegraphics[width=8.6cm]{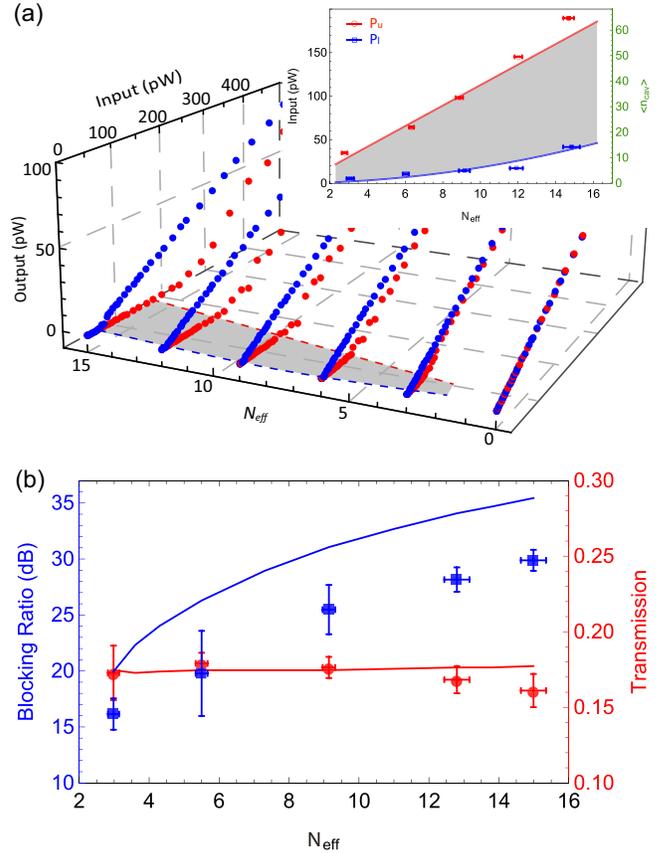}
\caption{\label{fig4} Performance of the nonlinear ONR for different effective atom numbers. (a) shows the measured bistability for the two counterpropagating light fields under different atom numbers. The blue and red points are for the light fields in modes $a$ and $b$, respectively. Each data point is the average of 20 measurements. The inset gives the map of the ONR working window associated with the corresponding atom numbers. The right vertical axis of the inset shows the corresponding mean photon number of the cavity for the forward light field. The solid curves and shaded area are the theoretical expectations, and the points are experimental data. (b) displays the average blocking ratios and transmissions of the ONR in the corresponding working window. Please see the SM \cite{N20} for extra data. The meaning of the points and solid curves is the same as that in Fig. \ref{fig3}.}
\end{figure}

The range of the ONR working window is bounded by upper and lower power $\text{P}_\text{u}$ and $\text{P}_\text{l}$, respectively, as indicated in Fig. \ref{fig2}, and the corresponding performance dramatically depends on the effective intracavity atom number $N_\text{eff}$. The bistabilities of the two light fields along opposite directions under different effective atom numbers $N_\text{eff}$ are measured and displayed in Fig. \ref{fig4}(a). As more atoms are involved and higher powers are requested to trigger the bistability for either side of the light fields, the ONR working window can be tuned by controlling the atom number. The inset of Fig. \ref{fig4}(a) gives the dependence of the upper and lower bounds of the ONR window on the atom number. It shows that if the atom number is low enough, for example, when $N_\text{eff}=3.0 \pm 0.2$, the ONR can work with a few intracavity photons. Of course, the working power of the ONR can also be high with a large number of atoms. Here, due to the limitation of the MOT and transferring system, the maximum $N_\text{eff}$ we could achieve was only $14.7\pm 0.3$. The effective number of atom could be increased further by improving the corresponding experimental setups.

The average transmission efficiencies and blocking ratios over the ONR working window for different atom numbers are shown in Fig. \ref{fig4}(b). As discussed before, the transmission efficiency for the transmitting light field is determined by impedance matching, which does not change with $N_\text{eff}$. The theoretical transmission efficiency therefore remains constant. The measured transmission efficiencies are in good agreement with the theory. The slight discrepancy is mainly due to the statistical errors of the measurement. The maximum blocking ratio of 30 dB can be achieved with $N_\text{eff}=14.7\pm 0.3$. The blocking ratios for the reverse light field decrease with decreasing $N_\text{eff}$ due to the weaker nonlinearity. However, because of the high nonlinearities, it is still higher than 15 dB even with $N_\text{eff}=3.0 \pm 0.2$ on a few-photon level. 

In summary, by using the high nonlinearity associated with atoms strongly coupled to an asymmetric optical cavity, we have experimentally demonstrated ONR on a few-photon level. Due to the high nonlinearity, the ONR can be observed with extremely low power at the $p$W level, corresponding to a few photons inside the cavity. The blocking ratio for the reverse-propagating light field is greater than 15 dB at this working power. The ONR working power window can be tuned by controlling the effective number of atoms, and the maximum blocking ratio can reach 30 dB. Comparing the nonlinear ONR with other conventional systems, we obtained two records of the lowest working power and the highest blocking ratio. The idea and results of ONR reported here can be easily integrated into optical chips as optical diodes by using a cavity QED system with chip-based Whisper-Gallery-mode cavities or fiber cavities. Our work opens up perspectives for nonmagnetic ONR on a few-photon level with quantum nonlinearities and has great potential for chip-based low-power photonic information technologies.

This work was supported by the National Key Research and Development Program of China (Grant No. 2017YFA0304502), the National Natural Science Foundation of China (Grant No. 11634008, 11864018, 11674203, 11574187, and 61227902), and the Fund for Shanxi "1331 Project" Key Subjects Construction.

\bibliography{Optical_Diode_bib}

\end{document}